# Experimental demonstration of an electrostatic orbital angular momentum sorter for electrons

Amir H. Tavabi*[1], Paolo Rosi*[2], Giulio Pozzi[1,7], Alberto Roncaglia[3], Stefano Frabboni[2], Enzo Rotunno[4], Peng-Han Lu[1], Robert Nijland[5], Peter Tiemeijer[5], Ebrahim Karimi[6], Rafal E. Dunin-Borkowski[1], Vincenzo Grillo[4]

1. Ernst Ruska-Centre for Microscopy and Spectroscopy with Electrons and Peter Grünberg Institute, Forschungszentrum Jülich, 52425 Jülich, Germany
2. Dipartimento FIM, Universitá di Modena e Reggio Emilia, 41125 Modena, Italy
3. Istituto di Microelettronica e Microsistemi-CNR, 40129 Bologna, Italy
4. Centro S3, Istituto di nanoscienze-CNR, 41125 Modena, Italy
5. Thermo Fisher Scientific, PO Box 80066, 5600 KA Eindhoven, The Netherlands
6. Department of Physics, University of Ottawa, Ottawa, Ontario K1N 6N5, Canada
7. Department of Physics and Astronomy, University of Bologna, 40127 Bologna, Italy

*Equal contributions.
Correspondence: vincenzo.grillo@cnr.it

**Abstract**

We report the first experimental demonstration of an electrostatic electron orbital angular momentum (OAM) sorter, which can be used to analyze the OAM states of electrons in a transmission electron microscope. We verify the sorter functionality for several electron beams possessing different superpositions of OAM states, and use it to record the electron beams OAM spectra. Our current electrostatic OAM sorter has an OAM resolution of 2 in the units of $\hbar$ - the reduced Planck constant. It is expected to increase the OAM resolution of the sorter to the optimal resolution of 1 in the future via fine control of the sorting phase elements.

**Introduction**

A modern transmission electron microscope can be used to characterize materials with sub-Å spatial resolution [1], to provide three-dimensional microstructural and compositional information [2] and to achieve an energy resolution of a few meV [3]. The introduction of spherical aberration correction [4-8] is a masterpiece of engineering, which requires the precise matching of the magnetic fields of hexapole (or quadrupole-octupole) lenses in both intensity and position.

Here, we realize an *electrostatic* "sorter" for electrons and demonstrate that it can be used to analyze components of the orbital angular momentum (OAM) of an electron beam [9-11]. Our OAM sorter also requires the perfect alignment and phase matching of several electron-optical components. In previous work, we reported the realization of such a device based on holographic phase elements that were fabricated from SiN [11]. In contrast, in the present study we demonstrate a completely electrostatic-field-based device, which can be retrofitted to existing electron microscopes.

The OAM sorter that we describe allows the measurement of the OAM state of an electron beam, for example after interaction with a sample [12][13][14]. Since the pioneering works that introduced the concept of electron vortex beams, it has become clear how to create electron probes that possess OAM [15][16][17][18]. However, the experimental measurement of this quantity has always been problematic. Methods that have been attempted include phase flattening [12] and diffraction through one or more apertures [13][14], which usually results in only a partial decomposition, which depends on the radius of the beam and is therefore not fully quantitative [19].

An OAM sorter has the advantages that: (i) it can be used to decouple radial and angular degrees of freedom; (ii) it provides a quantitative parallel spectrum of OAM; (iii) it is based on a unitary transformation that conserves beam intensity, thereby it is lossless and improves the measurement efficiency. Applications are foreseen in areas that include electron magnetic circular dichroism (EMCD) [20], structural biology , plasmonics [21], and the magnetic analysis of nanoparticles [11] . Most of these applications cannot be addressed easily using holographic solutions, which strongly reduce the intensity and quality of the electron beam.

In this work, we briefly describe the technical steps that are required to produce an electrostatic OAM sorter. We present preliminary results that demonstrate its successful operation, including the acquisition of OAM spectra in a transmission electron microscope.

**Experimental setup**

Sorter phase elements in an electron microscope are used to produce a conformal transformation of the electron beam from Cartesian to polar coordinates. The rotation, magnification and positioning between the elements must therefore be controlled perfectly. Fortunately, the simple addition of phase elements to replace pre-existing apertures and the smart use of deflectors can be used to produce the new configuration, while leaving the rest of the microscope almost unaltered.

Figure 1 shows the configuration of an FEI (Thermo Scientific) Titan-HOLO transmission electron microscope in Forschungszentrum Jülich, in which the positions of the phase elements and their predicted effect on the electron beam are illustrated. The microscope is operated at 300 keV and equipped with an X-FEG emitter. The sample is located in the standard position in the objective lens. The first sorter element S1 is located in the objective (OBJ) aperture plane, while the second sorter element is located in the selected area diffraction (SAD) aperture plane. In this way, the second sorter element S2 is located in the diffraction plane of the first sorter element, which is also conjugate to the sample plane. In this configuration, an OAM spectrum is produced when the microscope is set to diffraction mode. The Titan-HOLO microscope has two such SAD planes (SAD1 and SAD2); in the present study, we used the upper one.

For testing purposes, in order to generate electron beams with well-defined OAM states, we introduced a SiN hologram into the condenser aperture plane which was used as test objects similar to those described in a previous work [11]. The sorter elements S1 and S2 were introduced into the electron column in special customized aperture holders, which allow exchangeable microelectromechanical systems (MEMS) chips with 8 electrical contacts to be connected to external power supplies, in addition to hosting normal diaphragms.

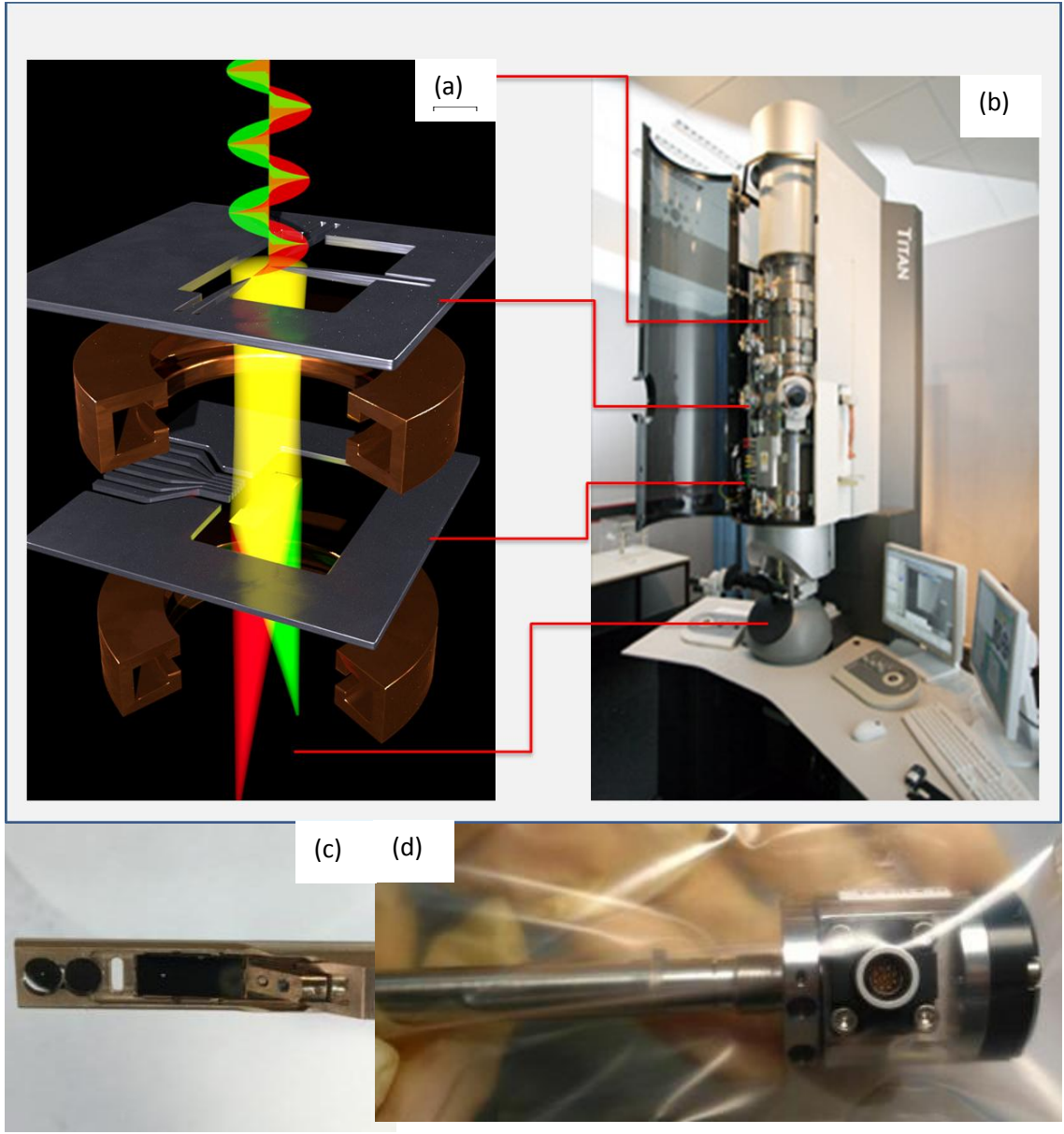

**Fig. 1** *(a) Concept of an electrostatic sorter and (b) the positions of its compnent elements in an FEI Titan transmission electron microscope in Forschungszentrum Jülich. (c) and (d) show details of the special aperture holder with contacts.*

According to the theory reported in [10], sorter element S1 should be a single, possibly very long, needle that is located in front of an electrostatic mirror. To a first approximation, such a needle can be modeled either as a straight line in a constant charge approximation or as an ellipsoid on the assumption of a fixed potential. As the latter approximation is more realistic (see the Supplementary Information), we shaped each of the needles into ellipsoids using focused ion beam (FIB) milling. We also added two lateral needles to the sorter element S1, based on our recent calculations [22], in order to compensate for the finite length of the wire, which would result in astigmatism and would affect the recorded OAM spectrum. Since each lens in the microscope introduces rotation, we placed particular attention on the orientations of the elements with respect to each other, as well as to the entrance of the Gatan Imaging Filter (GIF) spectrometer. Under standard working conditions, we measured a rotation angle of ∼-23° between the OBJ aperture and the first SAD aperture (SAD1) and of ∼26° between the OBJ aperture and the second SAD aperture (SAD2).

Scanning electron microscopy (SEM) images of an actual device are shown in Fig. 2. The MEMS chips were fabricated using optical lithography from a doped Si wafer and further shaped in three dimensions using FIB milling to produce the required electrode shapes. The corresponding schemes are also reported.

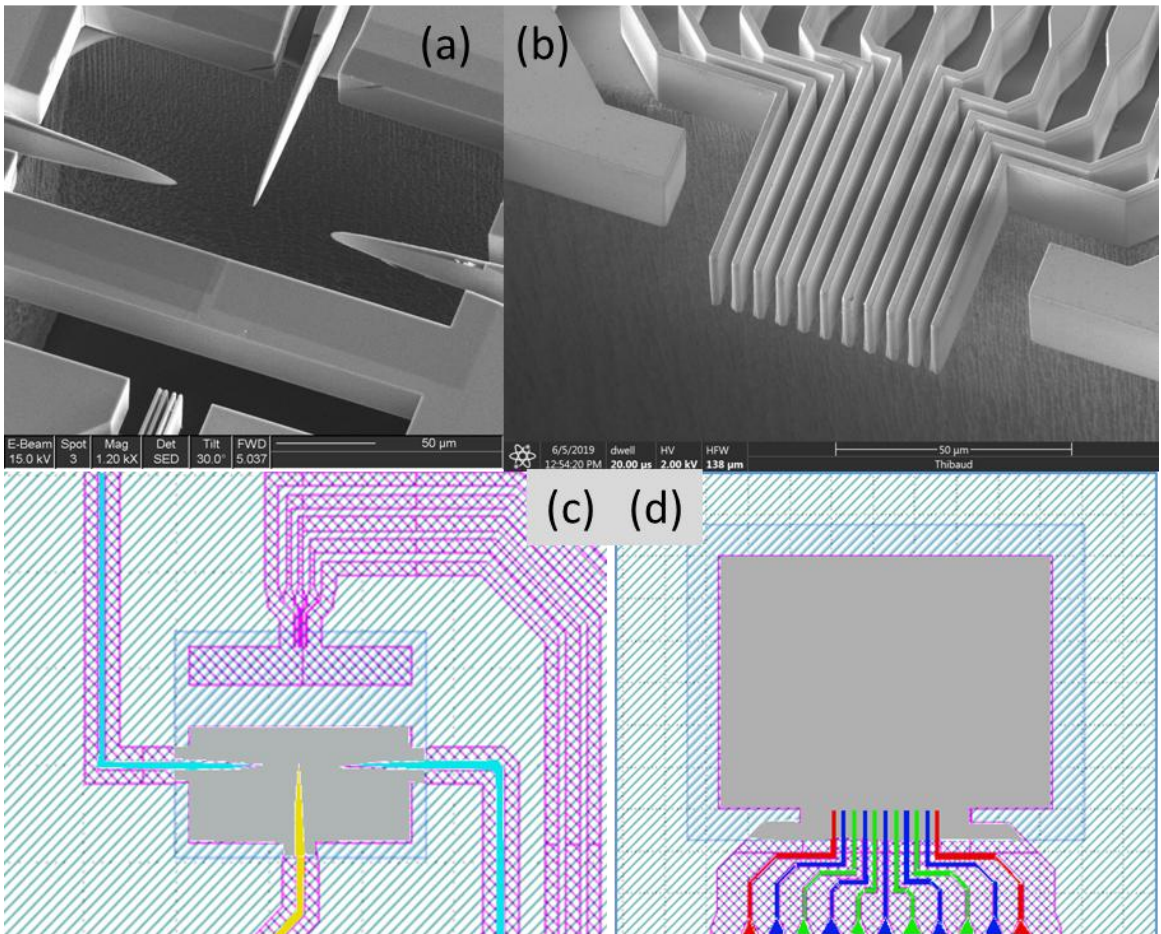

**Fig. 2** *(a, b) Scanning electron micrographs of the electrodes in electrostatic elements (a) S1 and (b) S2. (c, d) Corresponding design drawings. The color coding is consistent with that used in Table 1.*

**Sorting in practice**

In order to test the microscope in sorting mode, we used an electron probe, which was focused on the sample plane with a semi-convergence angle of 2.6 mrad. It has a diameter of ~20 µm in the OBJ aperture plane. The central needle of sorter element S1 was positioned mechanically at the center of the beam using the same motors as used for conventional apertures. In the image plane, the diffraction pattern of sorter 1 appeared as a rectangle, which was aligned approximately to the sorter 2 array. For a correct alignment the side of the rectangle must be exactly as large as 2 periods of the comb of electrodes of S2. Slight tuning of the corrector adapter lens (ADL) could then be used to fine-tune the rotation angle, while a variation in the voltage applied to sorter element S1 could be used to change the lateral extension of the rectangle. After an approximate mechanical alignment of the rectangular beam with sorter element S2, fine tuning was performed using the image shift coils. A switch of the lenses to diffraction mode then allowed the OAM spectrum to be viewed on the screen, or using the GIF camera to increase the magnification. In this mode, precise alignment of the beam shift was used to minimize the size of the diffraction pattern and to obtain a spectrum composed of a single line. The voltages that were applied to the electrodes for sorter elements 1 and 2 are given in Table 1.

Table1. Voltages applied to the electrodes for sorter elements 1 and 2.

|  | S1 central (yellow) | S1 lateral (light blue) | S2 lateral (red) | S2 odd (blue) | S2 even (green) |
|---|---|---|---|---|---|
| Applied Voltage (V) | -9.1 | 4.8 | -8 | -16 | 16 |
| Number of electrodes | 1 | 2 | 2 | 4 | 5 |

**Results**

As mentioned above, a special condenser aperture containing an in-line hologram that produces a combination of OAM states was used to calibrate the OAM spectrum and to estimate the OAM resolution. The first example is an electron vortex beam, which has an expected topological charge $\ell$ of 10. Although it was fabricated to take the form of a pure state vortex beam, imprecision in thickness control and the influence of the hologram on the electron amplitude introduced unintended harmonics into the OAM spectrum. In order to understand these additional features, we begin by establishing that, if the sorter element S1 is perfectly centered, the electron beam OAM state can be described in the form

$$|\psi\rangle = \sum_{n=-\infty}^{+\infty} a_{10 \cdot n} |10 \cdot n\rangle.$$

Any component in the final OAM spectrum that is not a multiple of 10 should therefore be an indication of poor OAM resolution. However, obstruction by the sorter tip also has a small effect. Figure 3 shows experimental images of the transformation of the electron wavefunction from the OBJ aperture plane to the SAD plane, as well as the final OAM spectrum. Due to the finite size of the S1 tip, part of the beam is obstructed. When it is transformed to Cartesian coordinates, only 9 of the 10 expected petals are visible and the obstruction introduces a degree of blurring into the OAM spectrum. However, simulations show that this blurring results in negligible broadening and the spurious intensities are below 5% in each neighboring channel.

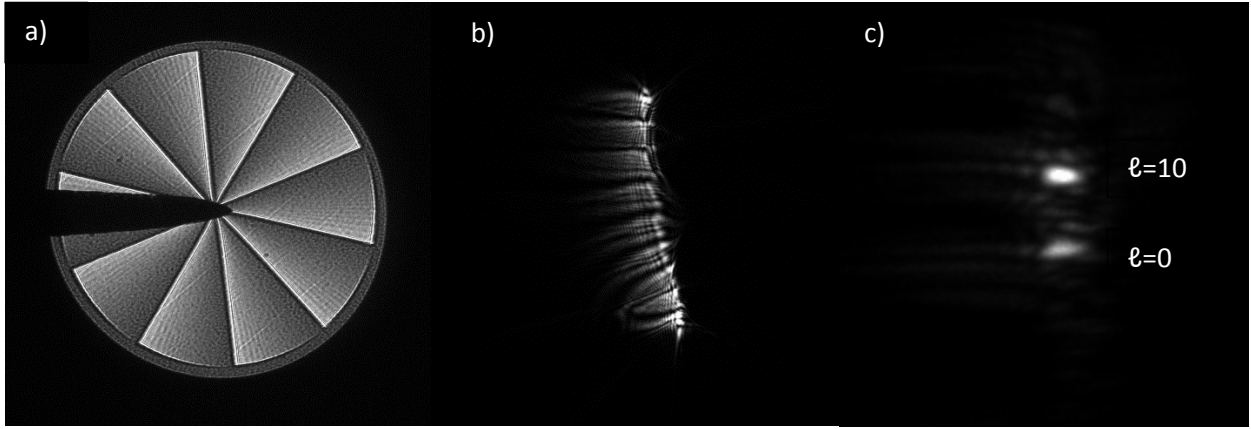

**Fig. 3** *Experimental images showing the evolution of the electron wavefunction for a nominal superposition of $|\ell = 10\rangle$ and $|\ell = 0\rangle$, starting from (a) its generation to (b) its conformal transformation to polar co-ordinates and (c) its transformation into an OAM spectrum.*

We also evaluated the OAM spectrum for several other in-line holograms and found an OAM resolution of ~2 (in unit of $\hbar$) as a full width at half maximum (FWHM) in the best case. A background and a set of spurious peaks are also present, as shown in Fig. 4. Appropriate software deconvolution is expected to produce sharp final peaks, as in the case of the holographic sorter.

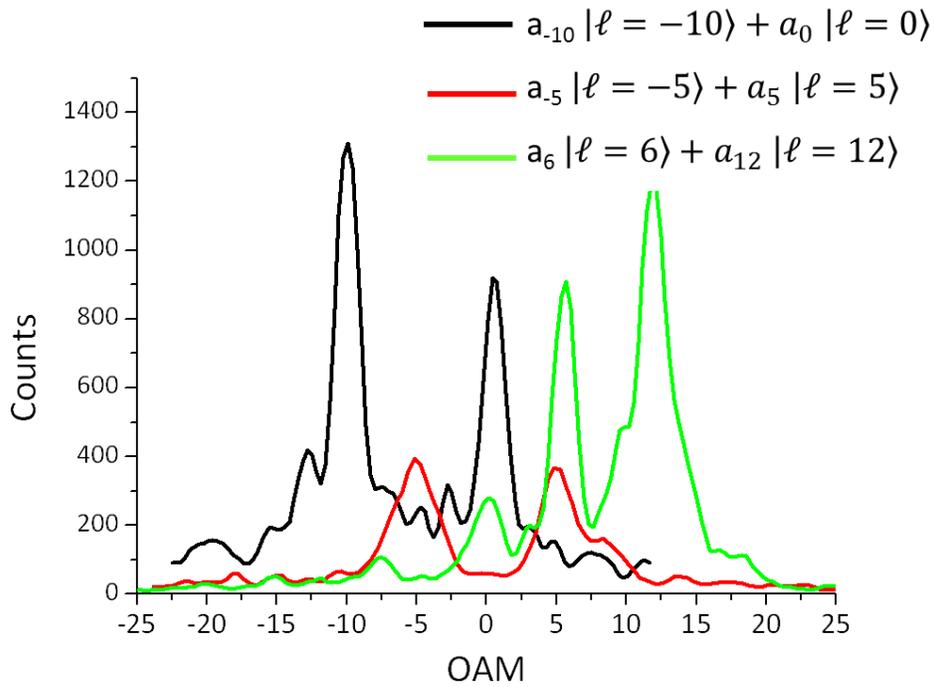

**Fig. 4** *Experimentally measured OAM spectrum for three test electron beams with the indicated nominal OAM compositions. The black curve corresponds to Fig. 3c.*

Our work demonstrates that a very simple procedure can be used to produce the phase alignment that is necessary to achieve OAM decomposition. The OAM resolution that we achieve is better than that obtained using the holographic approach, which demonstrated an OAM (FWHM) resolution value of $\Delta\ell \sim 2.5$ before off-line deconvolution of the point spread function. At the same time, the result is worse than expected from a simulation that is based on the measured phases of the sorting elements. (See S2 of the Supplementary Information). However, the phases were measured in the sample position using off-axis electron holography, where the voltage of the electrodes can be fine-tuned to obtain the desired phase. This is not possible as easily under real experimental conditions, when the phase elements are in the OBJ or SAD apertures.

In the rest of this paper, we provide a brief overview of the electrostatic parameters that we controlled, and which need to be refined further to obtain better resolution.

**Influence of astigmatism correction electrodes**

According to previously developed theory [22], two lateral needles should be placed on the sides of the main sorter element S1 needle at a voltage $V_L = -0.5\, V_C$. In order to test the working mode of the lateral needles, we systematically changed $V_L$. $V_C$ was also modified slightly, in order to retain the same S1 diffraction size.

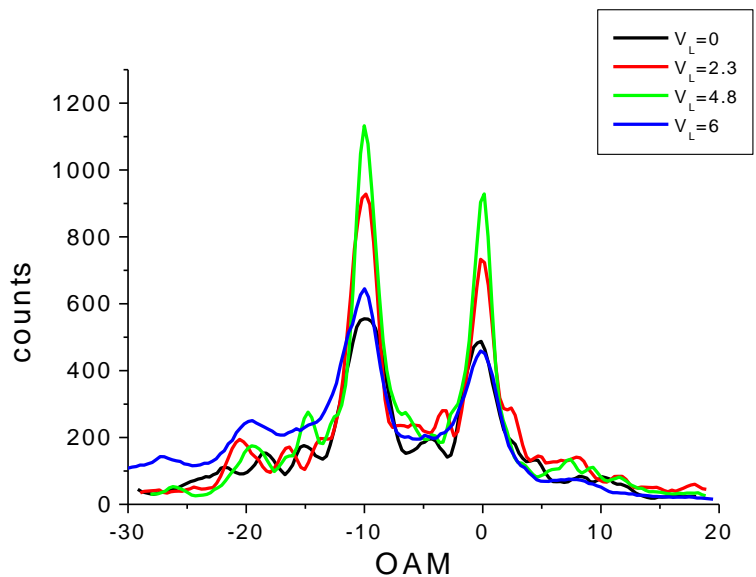

**Fig. 5** *Experimentally measured OAM spectrum for a superposition of $|\ell = 10\rangle$ and $|\ell = 0\rangle$ for different values of the voltage on the lateral electrodes. The starting voltage on the central tip was $V_C \sim -9.1\, V$.*

The results are shown in Fig. 5, which illustrates the fact that the optimal OAM spectrum is obtained when $V_L$ is very close to $-0.5\, V_C$, in addition to the importance of the lateral needles for astigmatism correction. This result is expected based on fixed charge theory and is confirmed using finite element simulations in the Supplementary Information. It is reassuring that the optimal voltage is predicted correctly using the analytical approach, as well as indicating that the electrodes have been shaped correctly.

**Phase of sorter 2**

When the sorter element S2 is located in the SAD1 plane, it is very difficult to measure its field. The current version of the sorter element S2 features 11 electrodes, which is close to the limit of our present fabrication technology and also of the number of contacts in our aperture holder. Accordingly, we also installed the sorter element S2 in the specimen plane, in order to measure its phase using off-axis electron holography. A representative phase measurement is shown in Fig. 6. It should be noted that there is a region of high spatial frequency oscillation in phase close to the electrodes that cannot be measured easily from this phase image. In contrast, far from the tips, the low spatial frequency behavior of the phase can be observed, and the general characteristics of the field can be inferred. One of the most prominent features is a lateral distortion of the phase on the right of the image, which results from the finite number of electrodes. Since the lateral size of the beam in the SAD plane will equal two times the spacing between the electrodes, we have some freedom where the position the beam with respect to the electrodes. We use this freedom to position the beam close to the central electrode, thus minimizing the influence of distortions due to the finite numbers of electrodes. The beam should also not be placed too close to any imperfection or asymmetry in electrode shape. Once the optimal distance from the electrodes has been identified, the voltage of each electrode can be selected accordingly.

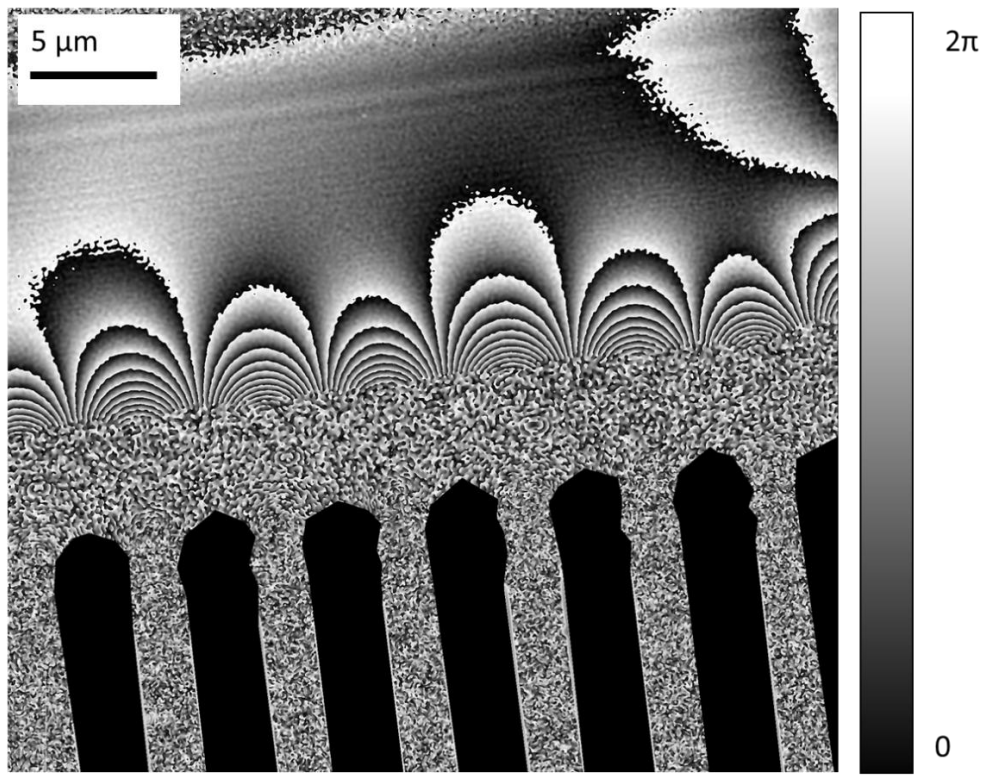

**Fig 6** *Phase distribution surrounding electrodes 3 to 9 in electrostatic sorter element S2 measured using off-axis electron holography. Distortions in the phase from that expected for an ideal periodic distribution result in part from the limited number of electrodes.*

## Conclusions

We have demonstrated the first experimental realization of an electrostatic electron OAM sorter. The sorter elements are fabricated using MEMS technology and are mounted in the OBJ and SAD apertures of a transmission electron microscope and are connected to external power supplies. An OAM spectrum can be obtained following alignment of the electrostatic phase elements in angle, position and magnification. By using in-line holograms, the OAM resolution value has been measured to be $\Delta\ell \sim 2$. It is improved by using astigmatism compensation in the S1 element and by tuning the boundary conditions in the S2 elements. Simulations suggest that it can be improved to the optimal value of $\Delta\ell = 1$ by tuning each electrode's voltage, as well as by improved accuracy in the positioning of both sorter elements.

## Acknowledgements

The authors acknowledge the support of the European Union's Horizon 2020 Research and Innovation Programme under Grant Agreement No 766970 Q-SORT (H2020-FETOPEN-1-2016-2017). E. K. acknowledges the support of Ontario Early Researcher Award (ERA), and Canada Research Chairs (CRC).

## References

[1] Y. Jiang, Z. Chen, Y. Han, P. Deb, H. Gao, S. Xie, P. Purohit, M. W. Tate, J. Park, S. M. Gruner, V. Elser and D. A. Muller, Nature **559**, 343-349 (2018).
[2] Y. Yang, C.-C. Chen, M. C. Scott, C. Ophus, R. Xu, A. Pryor, L. Wu, F. Sun, W. Theis, J. Zhou, M. Eisenbach, P. R. C. Kent, R. F. Sabirianov, H. Zeng, P. Ercius and J. Miao, Nature **542**, 75-79 (2017).
[3] O. L. Krivanek, T. C. Lovejoy, N. Dellby, T. Aoki, R. W. Carpenter, P. Rez, E. Soignard, J. Zhu, P. E. Batson, M. J. Lagos, R. F. Egerton and P. A. Crozier, Nature **514**, 209–212 (2014).
[4] M. Haider, S. Uhlemann, E. Schwan, H. Rose, B. Kabius and K. Urban, Nature **392**, 768 (1998).
[5]. O. Krivanek, N. Dellby and A. Lupini, Ultramicroscopy **78**, 1–11 (1999).
[6]. P. Hawkes, Phil. Trans. R. Soc. A **367**, 3637–3664 (2009).
[7]. O. Scherzer, Zeitschrift für Physik A Hadrons and Nuclei **101**, 593-603(1936).
[8] H. Rose, Journal of Electron Microscopy **58**, 77–85 (2009).
[9] G. C. Berkhout, M. P. Lavery J. Courtial, M.W. Beijersbergen, and M. J. Padgett Phys. Rev. Lett. **105**, 153601 (2010).
[10] B. McMorran, T.R. Harvey and M.P.J. Lavery, New Journal of Physics **19**, 023053 (2017).
[11] V. Grillo, A.H. Tavabi, F. Venturi, H. Larocque, R. Balboni, G.C.Gazzadi, S. Frabboni, P.H. Lu, E. Mafakheri, F. Bouchard, R. E.Dunin-Borkowski, R. W. Boyd, M. P. J. Lavery, M. J. Padgett, and E. Karimi Nature Comm. **8**, 15536 (2017).
[12] K. Saitoh, Y. Hasegawa, K. Hirakawa, N. Tanaka and M. Uchida, Physical Review Letters **111**, 074801 (2013).
[13] L. Clark, A. Bechè, G. Guzzinati and J. Verbeeck, Physical Review A **89**, 053818 (2014).
[14] G. Guzzinati, L. Clark, A. Béché and J. Verbeeck, Physical Review A **89**, 025803 (2014).
[15] K.Y.Bliokh, Y.P.Bliokh, S.Savel'ev, F.Nori, Physical Review Letters **99**(2007)190404.

[16] M. Uchida and A. Tonomura, Nature **464**, 737 (2010).
[17] J. Verbeeck, H. Tian and P. Schattschneider, Nature **467**, 301 (2010).
[18] B. J. McMorran, A. Agrawal, I. M. Anderson, A. A. Herzing, H. J. Lezec, J. J. McClelland, and J. Unguris, Science **331**, 192 (2011).
[19] H. Qassim, F. M. Miatto, J. P. Torres, M. J. Padgett, E. Karimi and R. W. Boyd, Journal of the Optical Society of America B **31**, A20 (2014)
[20] E. Rotunno, M. Zanfrognini, S. Frabboni, J. Rusz, R. E. Dunin Borkowski, E. Karimi and V. Grillo, arXiv:1905.08058.
[21] M. Zanfrognini, E. Rotunno, S. Frabboni, A. Sit, E. Karimi, U. Hohenester and V. Grillo, ACS Photonics **6**, 620-627 (2019)
[22] G. Pozzi, V. Grillo, P.-H. Lu, A. H. Tavabi , E. Karimi and R. E. Dunin-Borkowski, arXiv:1906.02344.

# Supplementary Information

## S1. MEMS design

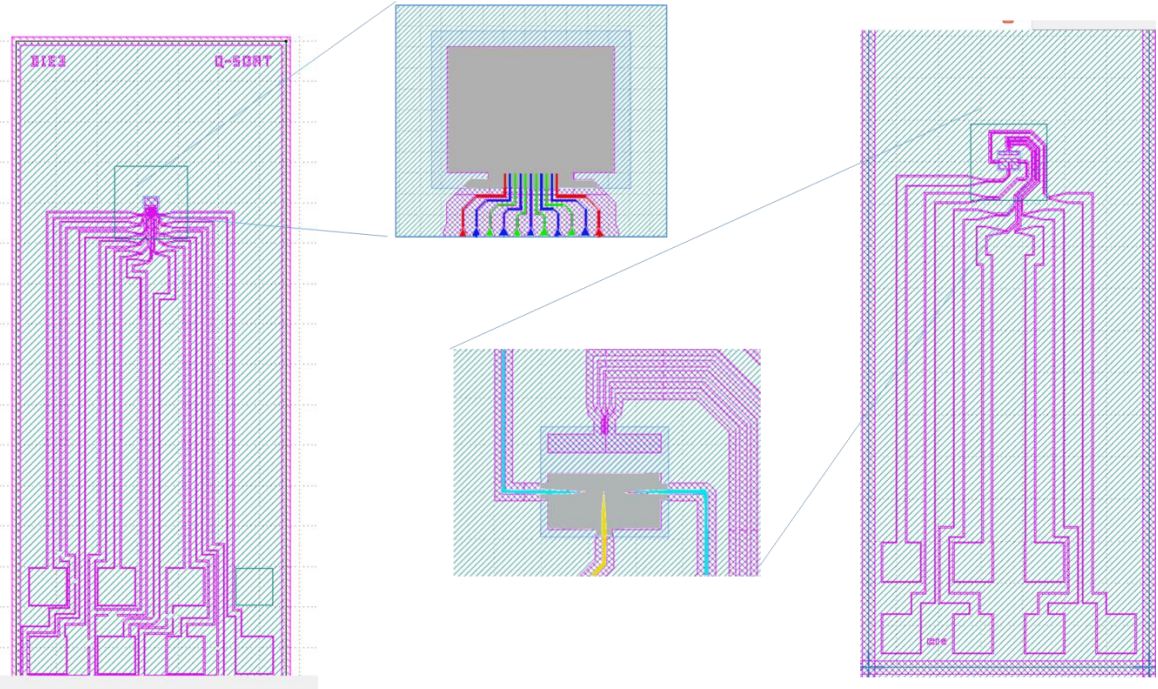

**Fig. S1** *Complete scheme of the MEMS for S1 and S2 from contacts to the active region.*

Figure s1 shows the design of the masks used for MEMS devices for the OAM sorter. The larger fields of view show the contact pads. The magnified images show the electrodes. Electrodes that are kept at the same potential are indicated with the same colors. In the text, the electrodes in sorter element S2 are referred to as "lateral" (red), "even" (blue), "odd" (green), while the electrodes in sorter element S1 are referred to as "central" (yellow) and "lateral" (bright blue). In sorter element S2, we are able to address 11 electrodes + ground using 8 contacts, while preserving the simplicity of the planar geometry.

## S2. Test of best expected performance

Figure S2 shows results of tests of early versions of the sorter phase elements, which had nominally worse specifications, using off-axis electron holography.

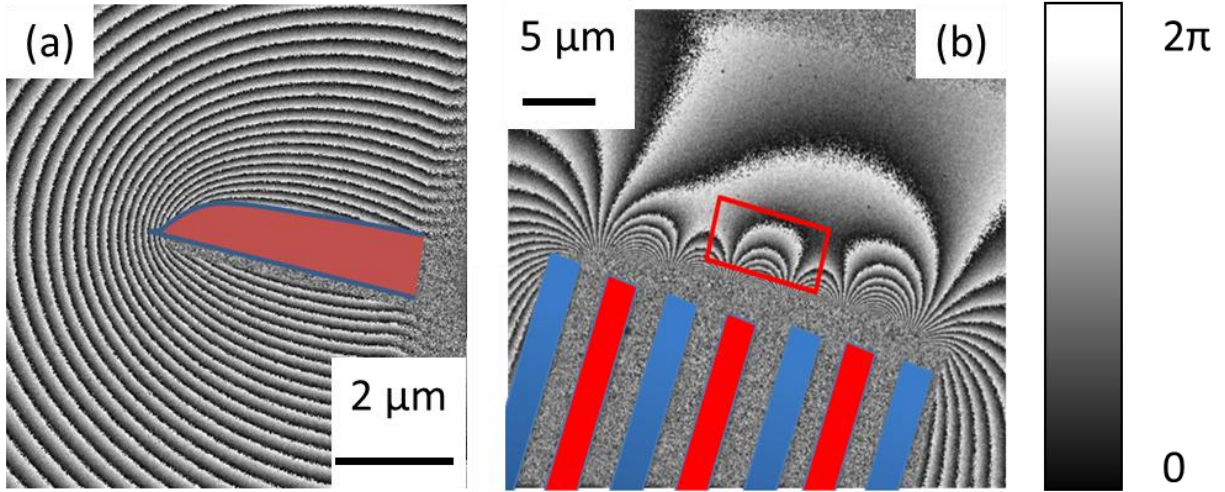

**Fig. S2** *Experimental phase distributions measured using off-axis electron holography for early prototypes of sorter elements S1 and S2.*

In order to simulate the experiment, we ensured that the diffraction of Sorter 1 was matched in size to the active area of Sorter 2. Using then an alignment algorithm of STEMCELL software we were able to find the most concentrate diffraction. This condition is what is also used in real experiments where the two elements are aligned by checking the actual OAM spectrum.

Mathematically, the alignment operation can be written in the form

$$\Psi(x,y) = \text{IFT}\left(\text{FT}(S1(x,y) \cdot P(x,y)) \cdot T_{\Delta u, \Delta v}(S2(u,v))\right),$$

where P is the probe to be sorted (in this case just a constant), FT and IFT represent forward and inverse Fourier Transforms, T is the translation operation necessary for alignment and the factors $S1(x,y) = \exp(i\varphi_1(x,y))$ and $S2(u,v) = \exp(i\varphi_2(u,v))$ represent the effects of the sorter phase elements.

Figure S3 shows (a) Sorter 1 diffraction calculated from experimental data in S2; (b) the Sorter 2 experimental phase distribution; (c) the resulting phase distribution after alignment; (d) the final spectrum. Since there was no structure in the original beam, the spectrum peak should correspond to the OAM = 0 condition. Unfortunately, Fig. S3c shows that the phase is not completely stationary, indicating that a part of the sorter is not perfectly aligned, possibly due to imperfections.

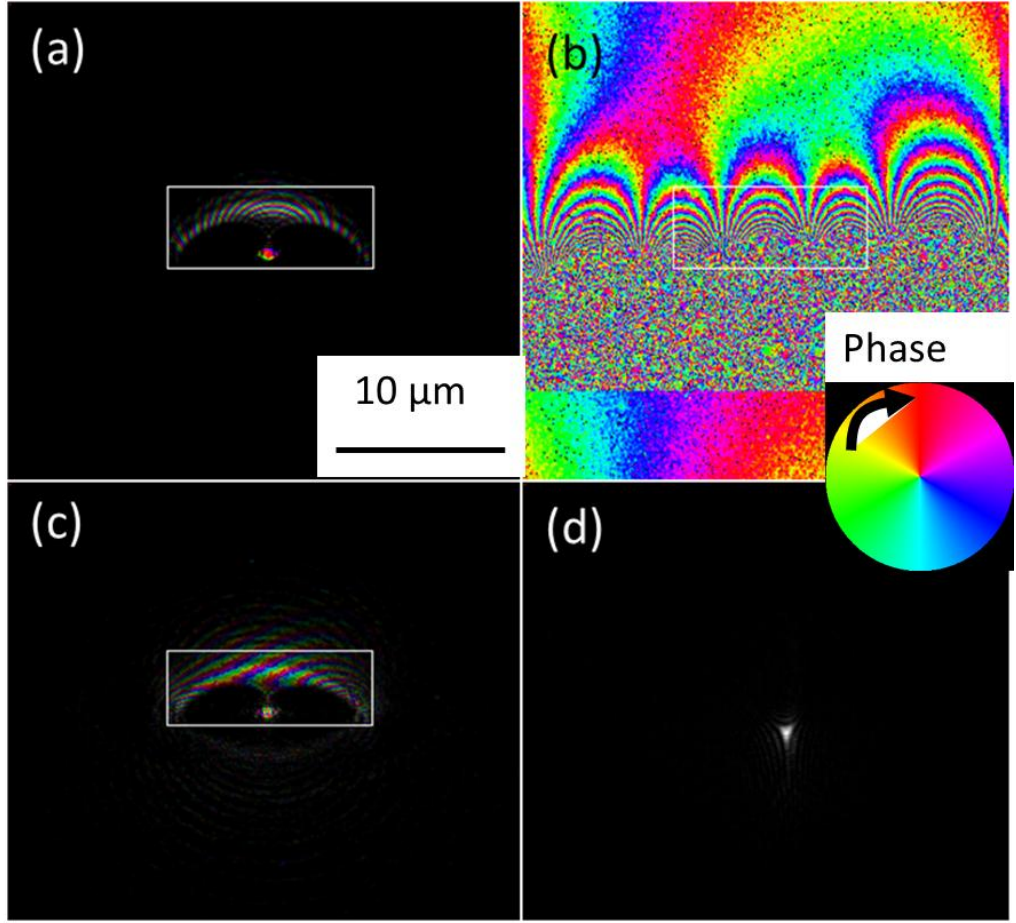

**Fig. S3** *Simulated sorting experiment using the experimental phase distributions of sorter elements 1 and 2. (a) The diffraction of sorter 1 is (b) superimposed on sorter 2 and after alignment the wave in (c) is diffracted to (d) a single line.*

Figure S4 illustrates the alignment procedure, going from an image of the beam in the sorter plane to the OAM spectrum, aiming at a smaller OAM spectrum.

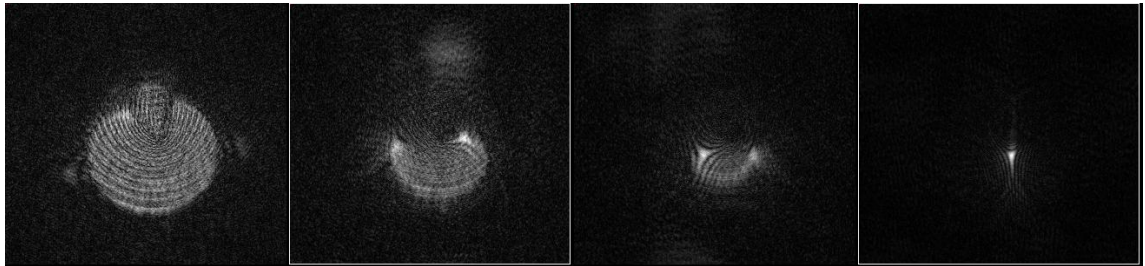

**Fig. S4** *Simulation of alignment procedure based on OAM spectrum inspection and on the experimental phase distributions of sorter elements 1 and 2. Varying the relative positions of sorter elements 1 and 2 is used to obtain a very localized peak in the ideal case.*

In order to assess the best obtainable resolution of the full sorter configuration, we used in our simulations a petal beam ($\ell = \pm 5$) and simulated iterative alignment of the experimental phase. The results are shown in Fig. S5. The alignment in orientation and magnification were performed visually (by looking at the OAM spectrum), as we found that the spectrum still allowed for very good separation of the peaks at $\ell = \pm 5$. This approach also allowed the OAM value standard deviation to be determined to be $\Delta\ell$=0.96.

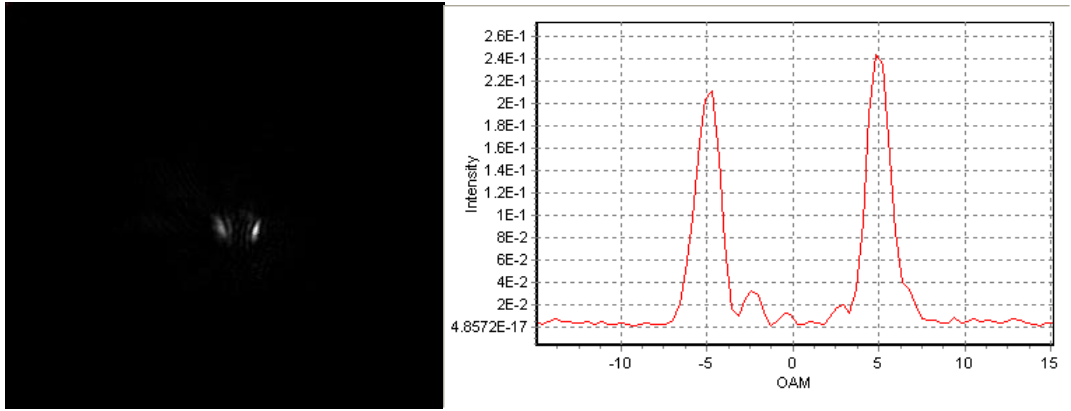

**Fig. S5** *Simulated spectrum based on experimental phases for a petal beam ($\ell = \pm 5$).*

The spectral lines are not perfectly straight, and side-lobes are evident. We anticipate that, through the control of each electrode and direct observation of the phase, the resolution can be further improved.

**S3. Tip shape**

The analytical model described if ref. [22] assumed a fixed charge on the needle in sorter element 1, astigmatism correction and the use of a mirror in front of the main needle. This model is too simplified. It is not completely realistic for the following reasons:

- We control the voltage and not the charge on the electrodes. The shapes of the electrodes can modify the charge distribution with respect to the ideal case.

- The large electrode in front of the tip has the same thickness as the tip itself. It cannot therefore be assumed to be a perfect charge mirror, as in the analytical model.

We used finite element simulations to assess these effects. COMSOL software can include a realistic description of the MEMS geometry. As this is a numerical method, we normally lose the predictive value of an analytical calculation. However, even before using numerical methods, we were already able to predict an interesting effect:

If the tip thickness is uniform, then the charge tends to concentrate towards the tip. The simplest model is to assume the charge as a uniform distribution (as in the ideal device), plus a concentrated charge at the tip. The main effect of such a charge is to add to the sorting potential a phase distribution (integral along z of the potential) of the form

$$\varphi = \varphi_0 \ln(|\bar{r} - \bar{r}_0|) ,$$

resulting in a logarithmic focusing of the beam before sorting.

We described the tip as (a) parallelepipeds, (b) elliptical cylinders or (c) ellipsoids. The relevant projected potential for each configuration is shown in Figs. S6 (d, e, f).

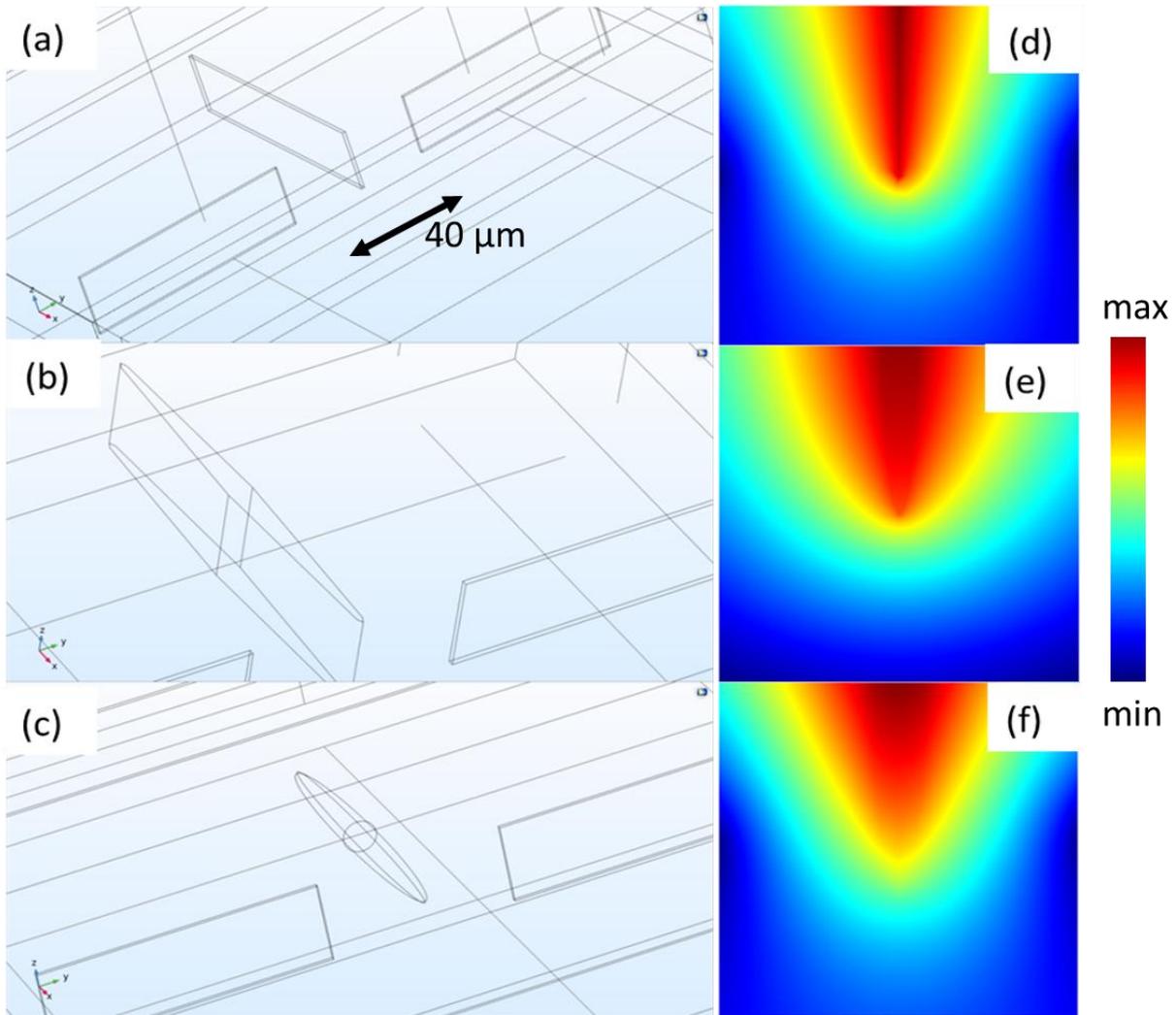

**Fig S6** *Geometry of the electrodes and total phase for sorter element 1 for different tip shapes: (a) parallelepipeds, (b) elliptical cylinders and (c) ellipsoids.(d,e,f) reports the projected potential for each case.*

The results of the calculation were exported as values of the potentials (1024 x 1024 x 100 points). This grid of potential values was then fed into software that converted them into 2D maps of integrated potential along z. The phase acquired by the electron at each point:

$$\phi(x,y) \propto \int_{-\infty}^{+\infty} V(x,y,z)dz$$

was approximated in the finite element calculation by an integral extending by about 100 µm above and below the device.

Unfortunately, the apparently smooth phase landscape resulting from these simulations is still not appropriate for checking the sorting properties. The finite element calculation intrinsically involves discontinuities, as the simulation scheme solves, for the electrostatic potential, the Laplacian equation separately in each sub-domain of the volume. It then connects the solutions in each domain continuously, resulting in a set of jumps in the derivatives between sub-domains. Since sorting is based on the local gradient of the sorter phase, these discontinuities in the derivatives affect the solution. We therefore took advantage of the fact that the phase must obey the relation $\nabla_{x,y}^2 \varphi = 0$ everywhere outside the tips and applied the following filtering procedure:

- Calculate $\nabla_{x,y}^2 \varphi$

- Apply a mask: where $\nabla^2 \varphi = \rho$, if $|\rho| < \varepsilon$ put $\rho = 0$

- Calculate the inverse Laplacian (in Fourier space it is just a multiplication).

This approach permits the generation of a smoother, and therefore more realistic, description of the sorting effect, with almost no discontinuities in the derivatives.

Assuming that the landscapes shown in Fig. S6 are calculated correctly, we need to assess which one is more similar to the theoretical image. For the sake of comparison, in Fig. S7 we plot the theoretical case. These parameters were selected by visual inspection to be close to the set of parameters reachable by the electrostatic device.

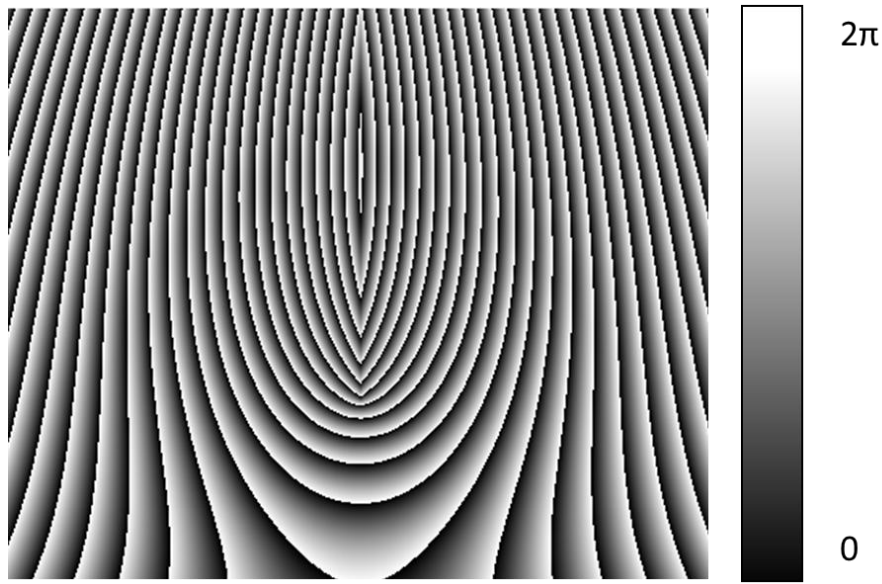

**Fig. S7** *Phase of an ideal sorter*

We note both similarities and differences between the realistic phase landscapes calculated by finite element simulations in Fig. S6 and the ideal sorter element 1 phase in Fig. S7.

Straight lateral parts are visible for all 3 configurations in Fig. S6, so it is difficult to understand which one is preferable. In order to understand these differences, we analyze the diffraction. Since the size of sorter element 2 is fixed by geometry, an increase in parameter A must be compensated by a different excitation of the lenses, in particular a smaller focal length of the overall lens system. This must be kept in mind during the experiment.

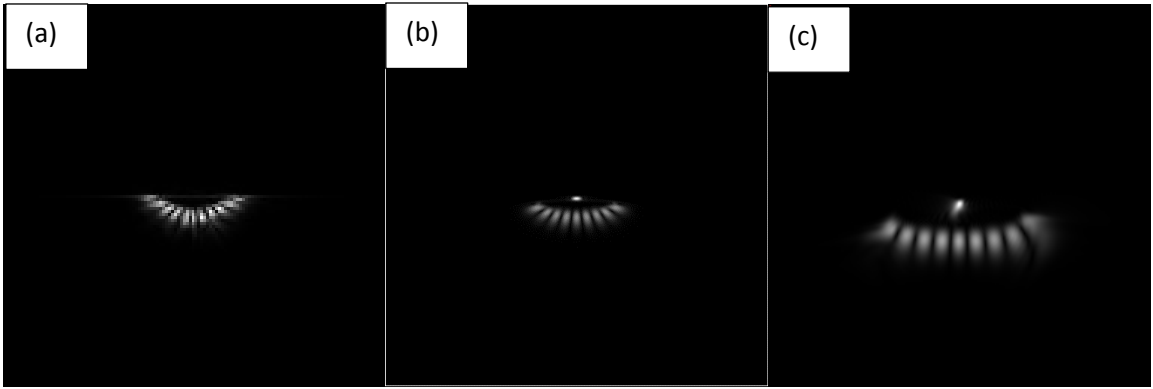

**Fig S8** *Diffraction of a petal beam ($\ell = \pm 5$) impinging on 3 different sorter 1 tips: (a) rectangular, (b) elliptic cylinder, (c) ellipsoid. Out of 10 petals, only 9 are visible in (b) and (c). The appearance of 11 petals in (a) probably results from spurious oscillations.*

For the simulation, we need to multiply the phase by a scale factor. We used a relatively large factor to test the effects of the different tips. The figure shows the effect of different tip shape on sorter 1 diffraction when a petal beam ($\ell = \pm 5$) is superimposed on the sorter. The tips produce different forms of bending of the diffraction. The ellipsoid, which is the most "correct" shape, produces the straightest line. As anticipated, the interpretation is that the logarithmic focusing term due to charge accumulation on the tip produces a sort of artificial focusing.

The tapering of the electrodes in the ellipsoidal electrodes seems to act as a better counterbalance of this charge accumulation.